# Dynamic Task Scheduling in Computing Cluster Environments

Ilias K. Savvas and M-Tahar Kechadi
Department of Computer Science,
University College Dublin, Belfield, Dublin 4, Ireland.

## Abstract

*In this study, a cluster-computing environment is employed as a computational platform. In order to increase the efficiency of the system, a dynamic task scheduling algorithm is proposed, which balances the load among the nodes of the cluster. The technique is dynamic, nonpreemptive, adaptive, and it uses a mixed centralised and decentralised policies. Based on the divide and conquer principle, the algorithm models the cluster as hyper-grids and then balances the load among them. Recursively, the hyper-grids of dimension $k$ are divided into grids of dimensions $k-1$, until the dimension is 1. Then, all the nodes of the cluster are almost equally loaded. The optimum dimension of the hyper-grid is chosen in order to achieve the best performance. The simulation results show the effective use of the algorithm. In addition, we determined the critical points (lower bounds) in which the algorithm can to be triggered.*

## 1 Introduction

A computational cluster (CC) can be defined as a set of independent computers (workstations, PCs, etc.) interconnected by a high-speed communication network such as Fast or Giga Ethernet [9]. The number of the participating processing elements or nodes can range from tens to hundreds [1, 3, 14, 17]. However, to fully exploit and effectively use any CC platform, resource management software must be provided to manage the complexity of different physical architectures for the user. This complexity arises in managing communication, synchronisation and scheduling a large number of tasks, in dealing with portability of libraries facilities used to parallelise/distribute user applications, and in many other related issues [13]. The scheduling of the submitted tasks to processing-nodes is a major concern with regard to performance and effective use of any CC. The problem of finding an optimal solution to the scheduling problem is NP-complete [16], where heuristic methods appear to be a suitable approach to solve this class of problems [15].

Scheduling tasks on computational clusters can be divided into two main classes: static and dynamic. In static scheduling, information regarding tasks' execution times and nodes' resources is assumed to be known beforehand. The assignment of tasks to processing nodes is done in such a way that the communication and the execution time of the whole application is minimised. While, dynamic scheduling techniques are based on the tasks assignment during their execution, taking into account over-loaded and under-loaded nodes, with the assumption that if the load among all nodes is balanced, then the overall execution time of the application is minimised. Here, the system has to decide, according to some centralised or distributed information, whether a task has to be transferred or not and to which node should be transferred [2, 6, 7, 8, 12].

In this paper, we propose a dynamic task scheduling technique based on "divide and conquer" principle. The proposed technique has two phases. During the first phase the network is mapped onto a hyper-grid. The second phase deals with the redistribution of tasks among the nodes by dividing the hyper-grid into hyper-grids of smaller dimension. Recursively, the load balanced hyper-grids of dimension $k$ are divided again into hyper-grids of $k-1$ dimension, until their dimensionality is equal to 1. The proposed technique is *dynamic*, *mixed*, *nonpreemptive*, and *adaptive* [5], and fully distributed. The transfer and placement decisions constitute the output of the technique.

The paper is organised as follows: In the next two sections the mathematical models of both network and tasks are presented. In section 4, we present our dynamic load scheduling technique. The section 5 discusses the performance issues of the technique and gives an example. The simulation results are given in section 6 and finally, concluding remarks and the future work are given in section 7.

## 2 System Model & Hyper-Grids

A Computational Cluster (CC) is a collection of independent processing-nodes interconnected by a network. We as-

sume that the following hold in a computational cluster:

- Each node $v_i$ is autonomous, has a full information on its own resources, and manages its work load.

- Each node $v_i$ has a processing power $\tau_i$ which represents the number of work units that can be executed per unit of time.

- The network uses a packet switched protocol and let $w$ be the size in bits of a packet, which is constant.

- The network's flow $b_{ij}$, which is the effective data rate in bits per second on the link that connect the nodes $v_i$ to $v_j$.

- The tasks are independent and can be executed on any node regardless its initial placement.

- There are two parameters associated with each task $t_i$: 1) the number of work units (in terms of computations) within the task ($\beta_i$), and 2) the number of packets required to transfer the task ($\mu_i$).

## 2.1 Hyper-Grids

Usually, a computational cluster has an irregular topology. This topology can be described by non-oriented graph $G(V, E)$, where $V$ represents the cluster nodes and $E$ the set of links between nodes. The first phase of this technique is to embed (map) the graph $G(V, E)$ into a multi-dimensional grid, called hyper-grid. The resulting grid is usually incomplete, in the sense that some of the links between neighbours and/or nodes are missing. The missing links and nodes are called *virtual links* and *virtual nodes* respectively. The second phase of our technique consists of recursively dividing the original hyper-grid into hyper-grids of smaller dimensions. The idea is to balance the load among the hyper-grids of the same dimensions starting from 1-dimension.
A $n$-dimensional grid ($G^n$) can be defined as a set of $n-1$-dimensional parallel hyper-grids as follows:

$$G^i = \{G_1^{i-1}, G_2^{i-1}, \ldots, G_{p_i}^{i-1}\} \quad i = n, \cdots, 1 \quad (1)$$

The hyper-grids of dimension 1 represent the nodes along one dimension (e.g. nodes connected by bus) and from the equation 1 one can deduce that the number of nodes of $G^n$ is $N = \prod_{i=1}^{n-1} p_i$. Therefore, we can define a hyper-grid recursively as follows:

**Definition 2.1.** An $n$-dimensional hyper-grid is a set of parallel $(n-1)$-dimensional hyper-grids. Zero-dimensional hyper-grids are the nodes of the system which are connected by links with the following properties:

- Links are either pair-wise vertical or parallel (considering that links which lie on the same line are the trivial case of parallel lines), and

- The length of links that connect direct neighbour nodes is the unity and it is constant.

Any node of the system can be represented as $v_{i_1, i_2, \cdots i_n}$. For clarity reason, let denote by $I$ the vector $[i_1, i_2, \cdots i_n]$. The dynamic task scheduling technique introduced in this paper has a phase that computes a one dimensional vector of loads of task hyper-grids. These loads are then balanced across the linear array of processor hyper-grids.

**Definition 2.2.** A hyper-grid load $W_x$ is the number of active tasks stored in nodes that are within hyper-grid of dimension $x$, called $x$-hypergrid.

This value is calculated by each processing-node for each hyper-grid that intersects it.

## 3 Task Scheduling and Allocation

The algorithm that inspired the techniques used in this study is the Positional Scan Load Balancing algorithm (PSLB) [11], which leads to a perfect load balanced system at a very reasonable time. The PSLB algorithm preserves the locality decomposition and it is based on the parallel prefix operation, or scan [4], which can be defined as follows:

**Definition 3.1.** The prefix sum operation $(+, A)$ takes the binary associative operator $+$, and an ordered set of n elements $A = \{a_0, a_1, \ldots, a_{n-1}\}$, returns the ordered set $\{0, a_0, (a_0+a_1), (a_0+a_1+a_2), \ldots, (a_0+a_1\ldots+a_{n-2})\}$.

### 3.1 PSLB Algorithm

The basic PSLB algorithm applies to 1-dimensional data-grids. PSLB algorithm is a very powerful dynamic load balancing algorithm, operating at the fine grain level. The generalisation of the algorithm to $n$-dimensional data-grids is also introduced in [11]. A brief description of the PSLB algorithm for grids of one dimension (line or bus) of system networks is given in *algorithm* 1.

### 3.2 PSTS Algorithm

In order to schedule more general applications (tasks) executing on irregular network topologies, we propose a technique based on PSLB, called the Positional Scan Task Scheduling (PSTS). PSTS approaches the task scheduling by applying the same technique as PSLB. PSTS uses the additive scan operation in order to find out the destination node for each work unit within each tasks. Firstly the algorithm indexes the work units (not the tasks), then uses the scan operator to collect information about the load in the system and processing powers, and finally for each node calculates locally the destination of each work unit. The key issue here is that, instead of considering a work unit as a basic processing unit, a task, consisting of many work units, is considered as a basic element. In other words, a task is a non-divisible load, however, the algorithm uses the work units to decide whether a task has to be migrated or not.

**Algorithm 1** PSLB Algorithm - Brief Description
1: Index the work units.
2: Use scan operator to collect information on the load in the system and on the processing powers.
3: Broadcast the collected information. (Each node knows how much work on the left hand side and the scan vector of the normalised processing powers)
4: For each node (in parallel): Calculate locally the destination of each work unit.
5: For each node (in parallel): Perform the migrations of the work units.

Let $T = \{t_1, t_2, \ldots, t_m\}$, represents the set of active tasks in the system, which consists of nodes $\{v_1, v_2, \ldots, v_n\}$. The total load in the system is

$$W = \sum_{i=1}^{m} \beta_i \quad (2)$$

As the system is heterogeneous, each node $v_i$ has different processing power, $\tau_i$, and is given as the number of work units can be executed per unit of time. So, the total processing power of the system is

$$\Pi = \sum_{i=1}^{n} \tau_i \quad (3)$$

In the algorithm, we utilise the normalised quantities. Let $\Pi = \sum_{i=1}^{n} \tau_i$ the total processing power of the system, the normalised processing of a node $v_i$, is $\gamma_i = \frac{\tau_i}{\Pi}$. Therefore, in a perfect load balanced system, the load of each node, according to the equations 2, and 3 is given by

$$W_i = W \frac{\tau_i}{\Pi} = W \gamma_i, \quad i \in \{1, 2, \ldots, n\} \quad (4)$$

The goal of the PSTS algorithm is to move each task $t_i$, from its current location $v_i$, to a node $v_j, v_j = F(t_i)$, so that the whole system is well balanced, and therefore the response time of different active tasks in the system is minimised; $R(t_i, v_i \to v_j)$.

PSTS algorithm, firstly calculates the exclusive additive scan of the work units of all tasks on the hyper-grids $G^1$ of dimension equal to 1.

$$S_{1r} = (+, L_{1r}), \quad r = 1, 2, \cdots, p_1 \quad (5)$$

where $L_{1x}$, is a vector of elements representing the number of work units of a node belonging to the hyper-grid $G_x^1$. Thus, $(+, L_{1r})$ is performed concurrently for all the hyper-grids of a dimension equal to one. In the same way, this operation is performed concurrently on hyper-grids of the same dimension, $p$.

$$S_{pr} = (+, L_{pr}), \quad r = 1, 2, \cdots, p_p \quad (6)$$

The resulting vector $S_{p*}$ represent the exclusive additive scans for all hyper-grids of the same dimension $p$. The total work load $W$ in the system is calculated by the operation $(+, S_{nr})$ on the $n$-dimensional hyper-grid. The scan operation is also used for determining the relative processing powers of the hyper-grids of different dimensions:

$$\lambda_{pr} = (+, \gamma_{pr}), \quad r = 1, 2, \cdots, p_p \quad (7)$$

This implies that each hyper-grid knows the normalised processing power of its hyper-nodes. The next step is to calculate the destination of each task within the hyper-grids of the same dimension. Each task $t_i$, according to its initial placement, is assigned to a node $v_I$, $v_I = F_{init}(t_i)$, and let $F(t_i) = v_J$. Then, the problem consists of calculating the index of the destination node $v_J$; $J = [j_1, j_2, \ldots, j_n]$. The calculations start from the highest dimension $n$ and continues until the dimension is 1. In order to balance the load in the system, algorithm calculates the least index $\lambda_n \leq i/W$. This means that the algorithm works according to the relative power of each hyper-grid and the sum of the work load of the entire system. After these scans each 1-dimensional hyper-grid knows whether is a "receiver" or a "sender". A receiver (resp. sender) means that a hyper-grid is under-loaded (resp. over-loaded). If, for instance, a 1-dimensional hyper-grid has to receive additional tasks then its own work consists of balancing its own workload according to the PSLB algorithm and just wait to receive more tasks which will go to the appropriate nodes. On the other hand, if it is over-loaded, then it uses the PSLB algorithm to balance its own workload (only for the tasks that have to remain in the hyper-grid), knowing that the hyper-grids of higher dimensions will balance the tasks among their elements, and therefore will migrate the extra tasks to their appropriate hyper-grids of lower dimension. This procedure guarantees that after its completion the entire system will be as close as possible to the perfect load balanced state. The description of the PSTS algorithm is given in algorithm 2.

## 4 PSTS Performance Model

Let $d$ denote the dimensionality of the grid. If $d = 1$ (bus topology) the number of communication steps needed is

**Algorithm 2** Positional Scan Task Scheduling Algorithm

1: **repeat**
2:    $r = 1$
3:    **for all** $r$-dimensional hyper-grids in parallel **do**
4:       $S_{rq} \leftarrow (+, L_{rq}), q = 1, 2, \ldots, p_r$
5:       $\lambda_{rq} \leftarrow (+, \tau_{pr}), q = 1, 2, \ldots, p_r$
6:    **end for**
7:    $W_r \leftarrow S^r + n_{\underbrace{i_1, p_1 - 1, i_3, \ldots, i_r}_{n-1}}$
8:    $\Pi_r \leftarrow \lambda_r + \tau_{\underbrace{i_1, p_1 - 1, i_3, \ldots, i_r}_{n-1}}$
9:    $r = r + 1$
10: **until** $(r = n - 1)$
11: $W \leftarrow \sum_{i=1}^{n-1} W_i + n_{\underbrace{p_2, p_1 - 1, p_3, \ldots, p_n}_{n-1}}$
12: $\Pi_r \leftarrow \sum_{i=1}^{n-1} \Pi_i + \tau_{\underbrace{p_2, p_1 - 1, p_3, \ldots, p_n}_{n-1}}$
13: **for all** 1-dimensional grids in parallel **do**
14:    Calculate, using the PSLB algorithm, if the 1-dimensional grid is a sender or a receiver
15:    **if** any 1-dimensional grid has to migrate tasks **then**
16:       Apply the PSLB algorithm for the destination 1-dimensional grid, and
17:       Migrate the tasks to the appropriate nodes.
18:    **else**
19:       Apply the PSLB algorithm for its own workload
20:    **end if**
21: **end for**
22: End.

$S^1_{comm} = 2(n-1)$, where $n$ is the number of the participating nodes, (see Figure 1). The number of computation steps is $S^1_{comp} = 2(n-1)$.

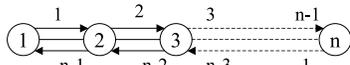

**Figure 1. d=1, comm. and comp. steps**

Let $p$ and $q$ be the costs in time units of a communication and a computation step respectively, then the total cost of the algorithm can be expressed as follows:

$$\begin{aligned} S^1_n &= S^1_{comm} + S^1_{comp} \\ &= 2(n-1)p + 2(n-1)q \\ &= 2(n-1)(p+q) \end{aligned} \quad (8)$$

For $d = 2$, the network topology is a grid consisting of $n_1$ lines and $n_2$ columns where $n = n_1 \cdot n_2$ (see Figure 2).

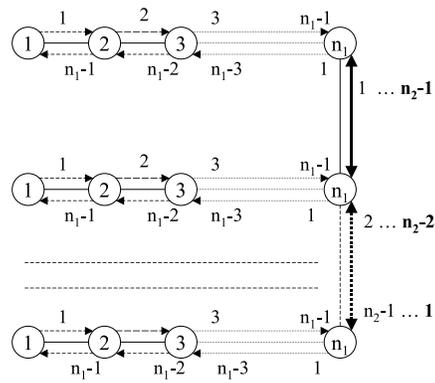

**Figure 2. d=2, comm. and comp. steps**

The number of communication and computation steps needed for each line (of $n_1$ nodes) is given by the equation 8; $2(n_1 - 1)p + 2(n_1 - 1)q$. This corresponds to balancing the load along each line of the grid (or hyper-grid of one dimension). Balancing the load along the columns can be done by performing the algorithm on hyper-grid of dimension 2 (by considering each line as a hyper-node of that hyper-grid). Therefore, the total cost for a 2-D grid is:

$$\begin{aligned} S^2 &= S^2_{comm} + S^2_{comp} \\ &= 2(n_1 - 1)p + 2(n_1 - 1)q \\ &\quad + 2(n_2 - 1)p + 2(n_2 - 1)q \end{aligned} \quad (9)$$

Finally, for $d = 3$ (figure 3), the total cost of computation and communication steps needed is:

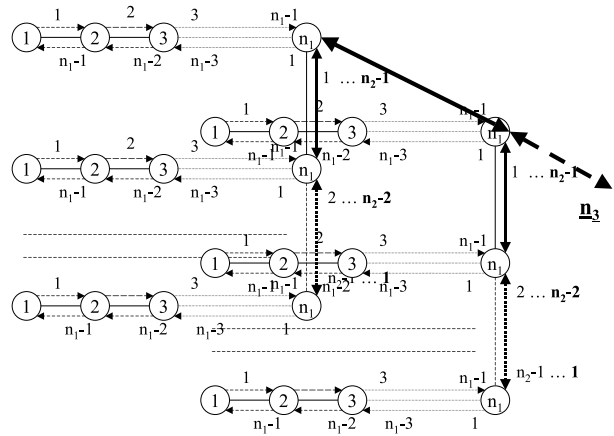

**Figure 3. d=3, comm. and comp. steps**

$S^3 = S^3_{comm} + S^3_{comp} =$
$2(n_1 - 1)p + 2(n_1 - 1)q + 2(n_2 - 1)p + 2(n_2 - 1)q + 2(n_3 - 1)p + 2(n_3 - 1)q$

or

$$\begin{aligned} S^3 &= S^3_{comm} + S^3_{comp} \\ &= 2(n_1 + n_2 + n_3 - 3)(p+q) \end{aligned} \quad (10)$$

and consequently for $d = k$

$$\begin{aligned} S^k &= S^k_{comm} + S^k_{comp} \\ &= 2(n_1 + n_2 + \cdots + n_k - k)(p+q) \end{aligned} \quad (11)$$

## 4.1 Embedding Irregular Network into $N$-D Grid

There are many ways of embedding an irregular network topology $G(V, E)$ into an $n$-D grid. The resulting grid is called incomplete and contains two types of nodes and links. Nodes (resp. links) which are mapped onto $V$ (resp. $E$) elements are called active nodes (resp. active links). The nodes (resp. links) which are not assigned to any element of $V$ (resp. $E$) are called virtual nodes (resp. virtual links). In order to ensure that the algorithm described above works on an incomplete grid, the virtual nodes are considered as active node with zero processing power. In the same way we can consider the virtual links as active links with zero bandwidth.

In order to minimise the cost of the PSTS algorithm on an incomplete grid, one need to minimise the number of virtual nodes or the dimension of the corresponding gird.

**Proposition 4.1.** *Consider a network $G(V, E)$ consisting of n nodes. The best performance of PSTS is achieved when $G(V, E)$ is mapped onto a $\lceil \log_2(n) \rceil$-D grid.*

*Proof.* According to the equations 8, 9, and 10 the communication and computational steps needed on an $\lceil \log_2(n) \rceil$-dimensional hyper-grid are:

$$S^{\log_2(n)} = S^{\lceil \log_2(n) \rceil}_{comm} + S^{\lceil \log_2(n) \rceil}_{comp}$$

$$= 2[\underbrace{(2 + 2 + \cdots + 2)}_{\log_2(n)} - \underbrace{(1 + 1 + \cdots + 1)}_{\log_2(n)}](p+q)$$

$$S^{\lceil \log_2(n) \rceil} = 2 \log_2(n)(p+q) \quad (12)$$

In order to prove that this dimension is optimal, the overhead produced (as in equation 12), has to be compared to the overhead produced on a grid of any other dimension that may embed the given network topology. Thus, by induction:

For $d = 1$, then it is obvious that:

$2 \log_2(n))(p+q) < 2(n-1)(p+q)$ since $\log_2(n) \leq n - 1, \forall n \geq 2$ and $\log_2(n) < n-1, \forall n > 3$.

For $d = 2$, using the equation 9:

$\log_2(n) < 2(n_1 + n_2 - 2) \Leftrightarrow \log_2(n_1 \cdot n_2) < 2(n_1 + n_2 - 2) \Leftrightarrow \log_2(n_1) + \log(n_2) < 2(n_1 + n_2 - 2), \forall n_1 \geq 2 \wedge n_2 \geq 2$, and analogous is the proof for $d = 3$.

If, for $d = k - 1$, it is true that:

$$\log_2(n) < 2[n_{x_1} + n_{x_2} + \cdots + n_{x_{k-1}} - (k-1)] \Leftrightarrow$$

$$\log_2(n) < 2(n_{x_1} + n_{x_2} + \cdots + n_{x_{k-1}} - k + 1) \quad (13)$$

Then, for $d = k$ it should be:

$$\log_2(n) < 2(n_{y_1} + n_{y_2} + \cdots + n_{y_{k-1}} + n_{y_k} - k) \quad (14)$$

But,
$2(n_{x_1} + n_{x_2} + \cdots + n_{x_{k-1}} - k + 1) \leq 2(n_{y_1} + n_{y_2} + \cdots + n_{y_{k-1}} + n_{y_k} - k)$
since at least one of $n_{y_1}, n_{y_2}, \ldots, n_{y_{k-1}}, n_{y_k}$ is less than at least one of the $n_{x_1}, n_{x_2}, \ldots, n_{x_{k-1}}$ in order to increase the dimension for at least one. $\square$

Thus, The highest dimension of hyper-grid that contains the initial topology $G(V, E)$ and gives the best performance using PSTS algorithm is

$$d_{optimum} = \lceil \log_2 n \rceil \quad (15)$$

## 4.2 Example

To help understanding the PSTS algorithm, we use a simple example of an irregular network and some tasks with their attributes. Consider a hyper-grid of dimension 2 and 18 nodes. It consists of three 1-D hyper-grids having six nodes each. The work load and the processing power per node and per hyper-grid, are given in table 1. For sake of clarity, we assume that each task consists of one work unit.

**Table 1. Proc. power and work-load / node.**

| Node | $v_{11}$ | $v_{12}$ | $v_{13}$ | $v_{14}$ | $v_{15}$ | $v_{16}$ | Total |
|---|---|---|---|---|---|---|---|
| Pr.Pow. | 3 | 4 | 5 | 2 | 1 | 5 | 20 |
| Tasks | 250 | 300 | 150 | 100 | 50 | 150 | 1000 |
| Node | $v_{21}$ | $v_{22}$ | $v_{23}$ | $v_{24}$ | $v_{25}$ | $v_{26}$ | Total |
| Pr.Pow. | 1 | 2 | 2 | 1 | 1 | 3 | 10 |
| Tasks | 200 | 300 | 100 | 400 | 300 | 700 | 2000 |
| Node | $v_{31}$ | $v_{32}$ | $v_{33}$ | $v_{34}$ | $v_{35}$ | $v_{36}$ | Total |
| Pr.Pow. | 5 | 1 | 4 | 2 | 6 | 2 | 20 |
| Tasks | 200 | 50 | 50 | 200 | 300 | 200 | 1000 |

The algorithm starts by calculating the exclusive scan vector for processing powers and the work load for all hyper-grids of dimension 1 (steps 2 to 6 as in algorithm 2). In steps 7 and 8, the rightmost nodes of each hyper-grid broadcast the total processing power, the total workload, and the vector of the normalised processing powers to the remaining elements of its hyper-grid. Table 2, illustrates 1-D hyper-grid calculations. When 1-D are completed, the algorithm performs the same steps for the 2-D hyper-grids as shown in table 3. As a result, each hyper-grid knows whether is a sender or a receiver. In this example, the hyper-grid $G_1^1$ is a receiver, $G_2^1$ is a sender, and $G_3^1$ is a receiver.

The application of PSLB algorithm on the two receivers, results in the redistribution of the work load as it is shown in table 4. The algorithm considers only the local work load of the hyper-grid, and balances the tasks among its nodes. On the contrary, when a hyper-grid is a sender, two major operations have to be done: 1) migration of the tasks, for which the index location is not in the local hyper-grid, to their new locations (communications between hyper-grids), 2) migration of some local tasks to their corresponding hyper-nodes, if their index location is different from the index of their current placement (communication between hyper-nodes of the same hyper-grid). The calculation of the index location of each task is possible as each hyper-node has a left neighbour and a right neighbour, knows how much work on its left side (calculated by scan operator), and the total workload in its hyper-grid (the up-right hyper-node broadcasts the total load once the scan is completed). In table 5, for instance, the nodes $v_{12}$ and $v_{16}$ have to send some tasks to $v_{13}$ and $v_{35}$ respectively.

**Table 2. PSTS: Steps $3 \to 8$, dim=1**

|            | $v_{11}$ | $v_{12}$ | $v_{13}$ | $v_{14}$ | $v_{15}$ | $v_{16}$ |
|------------|------|------|------|------|------|------|
|            | 0    | 3    | 7    | 12   | 14   | 15   |
|            | 0.15 | 0.2  | 0.25 | 0.1  | 0.05 | 0.25 |
| $\lambda_{1r}$ | 0    | 0.15 | 0.35 | 0.60 | 0.70 | 0.75 |
| $S_{1r}$   | 0    | 250  | 550  | 700  | 800  | 850  |
| $W_1$      | 1000 | 1000 | 1000 | 1000 | 1000 | 1000 |

**Table 3. PSTS: Steps $1 \to 10$, dim=2**

| Hyper-grid number | 1 | 2 | 3 | |
|---|---|---|---|---|
| $\tau$ | 20 | 10 | 20 | |
| Scan Proc.Power | 0 | 20 | 30 | 50 |
| Norm. Proc. Power | 0.40 | 0.20 | 0.40 | |
| $\lambda$ | 0 | 0.40 | 0.60 | 1 |
| $W_i$ | 1000 | 2000 | 1000 | |
| $S$ | 0 | 1000 | 3000 | 4000 |
| $W$ | 4000 | 4000 | 4000 | |

Finally, once all the operations (scan, send, receive, etc.) are terminated, each node of the system is affected a new load according to its processing power and the total load in the whole system. The PSLB algorithm, as one can notice, preserves the locality distribution. This means that data which are neighbours before the redistribution are more likely to stay neighbours after the redistribution. The advantage of locality distribution is that it can deal with divisible and indivisible load. For this application, each tasks is modelled as a set of work-units. But if part of the task has to migrate to another node, a decision has be made on whether the whole task has to migrate or not. Therefore, at the end of a redistribution, the system may not be perfectly balanced.

**Table 4. PSTS algorithm - "Sender" hyper-grid**

| Node | $v_{21}$ | $v_{22}$ | $v_{23}$ | $v_{24}$ | $v_{25}$ | $v_{26}$ |
|---|---|---|---|---|---|---|
| P.P. | 1 | 2 | 2 | 1 | 1 | 3 |
| W.L. | 200 | 300 | 100 | 400 | 300 | 700 |
| M. | 120 | 180 | 60 | 240 | ***180*** | ***420*** |
| S.M. | 0 | 120 | 300 | 360 | ***0*** | ***180*** |
| R.W.L. | 80 | 120 | 40 | 160 | 120 | 280 |
| S.R.W.L. | 0 | 80 | 200 | 240 | 400 | 520 |

**Table 5. PSTS algorithm - Migrating work load**

|   | $v_{11}$ | $v_{12}$ | $v_{13}$ | $v_{14}$ | $v_{15}$ | $v_{16}$ |
|---|---|---|---|---|---|---|
|   | 0 | 0.15 | 0.35 | 0.60 | 0.70 | 0.75 |
|   | $v_{21}$ | $v_{22}$ | $v_{23}$ | $v_{24}$ | $v_{25}$ | $v_{26}$ |
|   | 120 | 180 | 60 | 240 | ***180*** | ***420*** |
|   | 0 | 120 | 300 | 360 | ***0*** | ***180*** |
| $k$ |   | 100 |   |   |   | 200 |
|   |   | 0.37 |   |   |   | 0.63 |
|   |   | ($\to v_{13}$) |   |   |   | ($\to v_{35}$) |
|   | $v_{31}$ | $v_{32}$ | $v_{33}$ | $v_{34}$ | $v_{35}$ | $v_{36}$ |
|   | 0 | 0.25 | 0.30 | 0.50 | 0.60 | 0.90 |

## 5 Experimental Results

Any technique, having the goal of reducing the response time of tasks, hence, improving the overall CC system performance, introduces additional overhead in both computations and communications, due to its execution. The technique is considered to be worth implementing if its overhead is smaller than gained performance.

In this section, we evaluate the performance of the PSTS technique and validate the theoretical model previously presented. In addition, we want also to study some particular behaviour of the technique, mainly the crossover point with regard to the system state. A crossover point is a level of imbalance that the system can reach before triggering a load-balancing algorithm. The performance of the proposed technique are tested on both theoretically and experimentally on a cluster of 16 SUN Ultra SPARC IIi and PC clusters).

At the application level, the simulator parameters, taken under consideration, are the task distribution and size. The total number of tasks submitted and tested in the system is $m = 4,000$. The number of work-units (computation) and the number of packets (transfer) of the tasks were randomly chosen according to two different probability distributions at different levels: uniform and Poisson. For each distribution, the same number of tasks were tested with exactly the same attributes in order to observe the differences of the system's behaviour. Finally, in order to be as close as possible to realistic network situations, the tasks' arrival did not start at time zero and certainly they did not terminate after a finite amount of time [10]. At the system level the heterogeneity is expressed by three major factors: the node's processing power (normalised: $(1 \rightarrow 10)$, the number of participating nodes $(1 \rightarrow 64)$, and the bandwidth of the interconnection network.

Figure 4 shows the time taken by the PSTS algorithm for different sizes of CCs embedded into 1-D hyper-grids. Figure 5 shows PSTS overhead when it is applied to hyper-grids of higher dimensions. The overall time of the algorithm decreases as the number of the participating nodes increases since the same number of calculation are performed in parallel on an increasing number of nodes. In addition, the overhead produced on the higher dimensional hyper-grids is significantly less than that of one produced by 1-D hyper-grids because of the increasing number of steps performed in parallel. The algorithm is highly parallel.

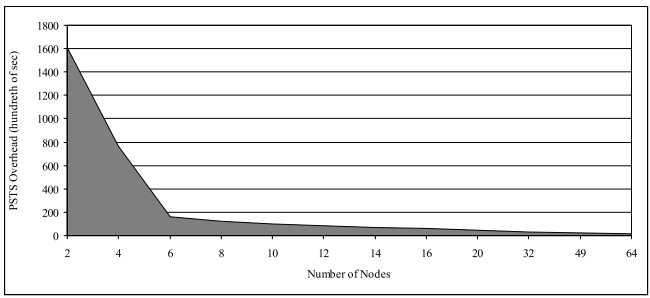

**Figure 4. Time taken for PSTS, dim=1**

Because of the overhead, the triggering phase of any dynamic scheduling/placement algorithm becomes crucial to the performance of the system. PSTS algorithm has the advantage of being highly parallel; its execution time is shared among the system nodes, but the crossover point at which its becomes beneficial to use PSTS algorithm remains to be determined. The crossover points for different number of nodes against the crossover points obtained by the use of higher dimensional hyper-grids are shown in table 6. One can see, again, that PSTS works better with hyper-grids of higher dimensions.

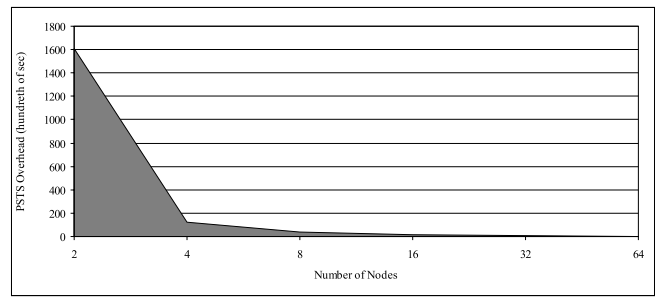

**Figure 5. Time taken for PSTS, dim>1**

**Table 6. Point at which load balancing becomes beneficial**

| Number of nodes | Crossover Point ($d = 1$) | Crossover Point ($d \geq 1$) |
|---|---|---|
| 2 | 1.0057 | 1.0057 ($d = 1$) |
| 4 | 0.6736 | 0.2058 ($d = 2$) |
| 8 | 0.4622 | 0.2979 ($d = 3$) |
| 16 | 2.0316 | 1.6069 ($d = 4$) |
| 32 | 2.7028 | 2.4228 ($d = 4$) |
| 64 | 3.0457 | 2.8701 ($d = 4$) |

The observed speedup gained by using PSTS for up to 64 nodes was calculated and shown in Figure 6. The speedup of PSTS decreases as the number of nodes increases, since the overall response time using the initial placements of the tasks is decreasing with the addition of more nodes to the system (the number of submitted tasks were kept constant). This is to illustrate the importance of calculating the crossover-point before triggering PSTS. The crossover-point depends mainly on the load of the system and evaluates the degree of the system imbalance.

We also observe the system behaviour when a new task arrives to the system. One wants to determine whether it is beneficial to apply the PSTS algorithm at any new arrival or not. This can be done by calculating the crossover point of the system with the new arrival. In table 7 the crossover points for different cluster sizes are presented. These crossover points are very low even for 1-D hyper-grid, and this suggest that PSTS could be applied after any new arrival.

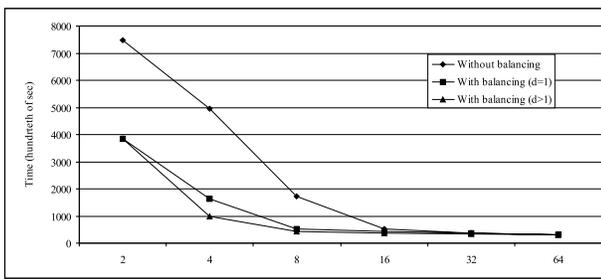

**Figure 6. Relative speedup of PSTS**

**Table 7. Point at which load balancing becomes beneficial for arrival of one new task.**

| Number of nodes | Crossover Point ($d = 1$) |
|---|---|
| 2 | 0.20333 |
| 4 | 0.15937 |
| 8 | 0.13593 |
| 16 | 0.12421 |
| 32 | 0.11835 |
| 64 | 0.11591 |

## 6  Conclusion

We proposed a new task scheduling mechanism for computing clusters, called PSTS. The PSTS algorithm is based on the PSLB algorithm, a pure dynamic load-balancing technique. We studied its cost and performance in both theoretically and experimentally. Is is shown that PSTS is highly parallel and efficient. It was shown from the crossover-points that PSTS can be used for very low imbalance systems. In addition, PSTS can be used as a scheduler for new tasks arriving to the system.

Further extension of this study will be at system level; for instance, include irregular networks where the routing problem will be the main issue. We would like also to extend this work to tasks which are node oriented and having dependencies between them.